\documentclass[aps,pre,twocolumn,groupedaddress,showpacs,floatfix]{revtex4}
\usepackage{hyperref}
\usepackage{graphicx}
\usepackage{amsmath}

\renewcommand{\vec}[1]{\mbox{\boldmath{$ #1 $}}}
\newcommand{\ten}[1]{\mbox{\boldmath{$ #1 $}}}
\begin{document}

\title{Projection operator formalism and entropy}
\author{E.A.J.F.\ Peters}
\email{e.a.j.f.peters@tue.nl}
\affiliation{Dept.\ of Chemical Engineering\\
Technische Universiteit Eindhoven\\
P.O. Box 513\\
5600 MB Eindhoven\\
The Netherlands}
\date{\today}

\begin{abstract}
The entropy definition is deduced by means of (re)deriving the generalized non-linear Langevin equation using Zwanzig projector operator formalism.
It is shown to be necessarily related to an invariant measure which, in classical mechanics, can always be taken to be the Liouville measure.
It is not true that one is free to choose a ``relevant'' probability density independently as is done in other flavors of projection operator formalism.
This observation induces an entropy expression which is valid also outside the thermodynamic limit and in far from equilibrium situations.
The Zwanzig projection operator formalism therefore gives a deductive derivation of non-equilibrium, and equilibrium, thermodynamics.
The entropy definition found is closely related to the (generalized) microcanonical Boltzmann-Planck definition but with some subtle differences.
No ``shell thickness'' arguments are needed, nor desirable, for a rigorous definition.
The entropy expression depends on the choice of macroscopic variables and does not exactly transform as a scalar quantity.
The relation with expressions used in the GENERIC formalism are discussed.
\end{abstract}


\maketitle

\section{Introduction}

The classical, Boltzmann-Planck, definition of entropy is the logarithm of the number of microstates corresponding to a macroscopic state of a system (times the Boltzmann constant $k_B$).
Within classical mechanics this definition causes some fundamental problems since the number of microstates is not countable.
The common resolution is to define a unit volume of microscopic phase space.
Within classical mechanics the motivation why this is reasonable is found in Liouville's theorem (incompressibility of phase space velocity).
To quantify what the unit of phase space should be one usually resorts to quantum mechanics, heuristically, to the uncertainty relation.
A unit volume in microscopic phase space, specified by positions and momenta of all $N$ particles, is according to this reasoning proportional to $\hbar^{3N}$.
In the setting of the microcanonical, iso-energy, ensemble this does not resolve the issue fully, since the iso-energy surface has a zero Liouville measure.
In this case a finite shell thickness is usually assumed.

Several reasonings are encountered to motivate this thickness.
The first is to refer, again, to quantum mechanics and the uncertainty relation.
A second reasoning is that the thickness of the shell region is somehow set by uniformity of fluctuations.
The validity of both reasoning can be debated especially because, within the classical setting, in a closed system there are no fluctuations of the total energy.
A third, well founded, reasoning is that the thickness is irrelevant for the case of the thermodynamic limit.
In this view entropy is only fully unambiguously defined in this limit.
Indeed the proof that the extensive entropy expression in this limit is independent of the shell thickness can be found in classical monographs \cite{Rue69, Lan73}.

In the current paper we will show that the definition that follows from the Zwanzig projection-operator formalism \cite{Zwa61} that leads to a generalized (non-linear) Langevin equation gives rise to an entropy definition that does need no reference to a shell thickness.
Furthermore it can be defined for any set of macroscopic variables.
In the case of non-conserved quantities, the shell argument is hard to defend (since an ensemble does not remain within such a shell).
One does not have to worry about that since the non-shell definition is the ``real'' entropy definition.

The usual reasoning of why entropy is an important quantity has to do with ergodicity-like arguments, i.e., sampling large parts of microscopic phase space by ``trajectories''.
In other words there is a connection to dynamics.
A formal mathematical tool to discuss the issue of the connection between microscopic dynamics and thermodynamics is projection operator formalism, \cite{Zwa01, Gra82}.
It is a method for decomposing equations of motion.
So, in principle it gives a different representation of an already known exact equation.
Its use is that it may point the way toward good modeling assumptions and toward well-chosen approximations.
When defining macroscopic variables to describe a system it can be used to ``project'' the microscopic dynamics onto the macroscopic phase space.

One might expect that, as a by-product of this procedure, the entropy definition arises.
Usually, this is not the way the projection operator is constructed.
In the conventional practice one is free to choose macroscopic variables as well as a ``relevant'' probability distribution \cite{Gra82}.
The relevant probability distribution follows from independent, statistical mechanics, reasoning.

There are many flavors of projection operations (Robertson, Mori, Zwanzig, Kawasaki etc.).
The reason is that one has quite a lot of freedom to define projection operators.
The common methodology is to make use of a Hilbert-space description of a system, e.g., in \cite{Nor75}.
One basic ingredient of creating such a Hilbert space is defining an inner-product.
Using this inner-product one can then construct projection operators.
As an input for defining the inner-product one uses the equilibrium distribution function of a system.
Since the equilibrium distribution is a result derived from statistical mechanics, with an implicit entropy definition, the hope to find an entropy definition from first principles is lost in this case.

An often used flavor of projection operator formalism, originated in the work of Mori \cite{Mor65}, results in generalized linear Langevin equations.
In this linear case, the system is characterized by the expectation values (or rather ensemble averages) of a set of macroscopic variables.
The resulting equations are useful only near thermodynamic equilibrium.
The reason is that expectation values of non-linear functions of the variables (higher moments), that are also likely to relax slowly, are not part of the set of macroscopic variables.
These non-linear functions of the macroscopic variables can ``hide'' in orthogonal subspaces of the Hilbert space and are therefore not filtered out by the projection operator.
These slow relaxation time scales show up as slowly decaying memory functions.
For useful approximation further from equilibrium, using this Mori approach, also higher order moments and cross-correlations (mode-couplings) have to be taken into account (see \cite{Zwa01} for a basic explanation).

A second flavor of projection operator formalism is the one introduced earlier by Zwanzig \cite{Zwa61}.
In this case the system is characterized by a full set of macroscopic variables, and not only the expectation values of the first moments.
This is much more restrictive on the choice of (reasonable) projection operators \footnote{This approach can be cast in the same formalism as the previous case by considering expectation values of delta-functions \cite{Nor75, Gra82}.}.
It produces a non-linear generalized Langevin equation.
The reason that, in the current paper, this flavor of the projection operator method is used is because, after making approximations, it has the potential to be valid also far from equilibrium.
Both the Mori equations and the Zwanzig equations are formally exact.
The Zwanzig approach, however, allows for better approximations in the non-equilibrium situations.
It is therefore probably a better starting point to develop non-equilibrium thermodynamics.

Also when applying formal projection operator techniques in the non-linear case one can, in principle, use any ``relevant'' distribution to define the projection operators.
The subtlety we want to point out here is that only specific properties of the distribution, namely, invariance with respect to the microscopic dynamics, allow one to derive the non-linear Langevin equation of a useful form.
In classical mechanics the Liouville measure is always invariant.
This, therefore is the most obvious choice to use, and defines entropy.
Note that this choice is not restrictive since the development of the equation is still strictly a formal decomposition.
No modeling assumptions are needed to arrive at this result. The fact that one is quite restricted in the choice of a sensible ``relevant'' distribution, in fact that it is necessarily an invariant distribution (or, rather, measure), is the main observation of this paper.

In early accounts of projection-operator formalism by Zwanzig \cite{Zwa61} he uses the same microcanonical distribution that follows from the arguments presented here.
And sometimes it is still applied, e.g., in \cite{Ott98}.
In later derivations, especially those that make use of a Hilbert space, e.g., \cite{Zwa01, Gra82}, a freedom to impose an equilibrium or relevant distribution is suggested.

In this paper we will circumvent, as much as possible mathematical constructs such as densities in phase space and vectors in Hilbert space.
The underlying philosophy is that an ensemble is a mathematical construct that should be used with caution.
A system is in ``reality'' only in one microstate.
By using a description in terms of densities one is tempted to see this as something ``real'', e.g., to view the ensemble as the state of a system.
By making approximations on the level of phase space densities, or on Hilbert space vectors, one can construct objects that have little to do with reality.
The fact that projection operator formalism has so many flavors is mainly due to choices made at this abstract level.
To stay as close to reality as possible we will try to remain at the level of equations of observables (quantities) and not ensembles or vectors in Hilbert space.

For the sake of completeness this paper will start out with giving the derivation of the generalized non-linear Langevin equation.
The derivation will put emphasis on the role of the invariant measure and appearance of an entropy related to this.
The reason is that an important goal of the current paper is to present entropy as something that arises naturally in the course of deriving the non-linear generalized Langevin equation.
One does not need independent statistical mechanical reasoning to pose a ``relevant'' probability distribution.
One needs no qualitative arguments, or connections to quantum mechanics, to justify the form of the entropy expression.

The definition of the entropy that arises is very similar to the Boltzmann-Planck-Einstein definitions, but there is a subtle difference.
In the definition found here $\exp[S(X)]$ is the number {\em density} of microstates corresponding to a macrostate.
So, when one considers a volume $\delta V_X$ as a small volume around macrostate $X$ then, in sloppy notation,
\begin{equation}
    \delta V_\Gamma = \exp[S(X)] \, \delta V_X
    \label{eq:phase_space_volume}
\end{equation}
is the volume (Liouville measure) of the corresponding region in microscopic phase space.
In this paper the Boltzmann constant $k_B$ will be put to $1$.
The entropy that arises in the ``Einstein distribution'' is exactly the entropy as defined here.

This definition raises several questions.
Firstly, entropy seems badly defined since the ratio of volumes is not dimensionless, so one is taking the logarithm of a dimensional number.
Therefore entropy transforms strangely upon change of dimensions (a constant should be added).
Related to this, entropy is not a scalar quantity.
Upon a coordinate transformation of the macroscopic space, entropy changes in a non-trivial way.
We will argue that this is just the way it is and everything works out fine.

The projection operator formalism suggests that one can define an entropy for any set of macroscopic variables.
A modern formalism for non-equilibrium thermodynamics, very much related to the current approach, is the GENERIC formalism \cite{Grm97,Ott97_2,Ott05}.
This formalism should be derivable from the generalized non-linear Langevin equation after some controlled modeling assumptions.
An attempt to proof the formalism on the basis of projection operator formalism can be found in \cite{Ott98, Pab01}.
There the same definition of the entropy is given as in Eq.~\eqref{eq:phase_space_volume}.
However, to avoid the conclusion that entropy does not behave as a scalar upon coordinate transformation, a preferred coordinate system is introduced from the macroscopic state space in \cite[p. 228]{Ott05}.
We will argue it is not necessary to introduce a preferred coordinate system.

The structure of GENERIC is richer than that found by means of the projection operator formalism alone.
There is a Poisson structure for the reversible part, and two extra degeneracy conditions.
In \cite{Ott98,Pab01,Ott05} this extra structure could not be proved, but was argued to be very likely.
The degeneracy conditions can be directly derived from projection operator formalism, but the Poisson structure can not.
This will be discussed.

The entropy expression directly follows from the non-linear Langevin equation before any modeling assumptions are made.
The usual derivation, in equilibrium thermodynamics, of entropy expressions use ergodicity arguments.
Since we derive entropy in a dynamical setting ergodicity is never strictly obeyed.
This raises questions about the relation between entropy and ergodicity. This will be discussed near the end of the paper.
The approach outlined here could be a starting point of applying non-equilibrium thermodynamics to systems outside the linear regime or outside the thermodynamic limit, i.e., for small systems.

\section{The nonlinear Langevin equation}
\label{section:Langevin}

The derivation of the nonlinear Langevin equation using projection operator formalism can be found in many standard texts and papers \cite{Zwa01, Gra82}.
Nevertheless, we will give a straightforward derivation here.
The reason for giving it is to make this paper self-contained and to be able to point out specifically where, and how, the notion of entropy enters.

This derivation of the non-linear Langevin equation will start out by closely following that of Kawasaki \cite{Kaw73}.
We will use $\Gamma$ to denote a point in microscopic phase space and $X$ for a point in macroscopic phase space.
To every point in microscopic phase space $\Gamma$ one point $X(\Gamma)$ corresponds.
This relation is generally not invertible.
A point in macroscopic phase space corresponds to a whole subspace in microscopic phase space.

For any quantity $A(\Gamma)$ the time development is described by means of a Liouville operator $\mathcal{L}$, formally,
\begin{equation}
  A_t = \exp[i \mathcal{L} t] A_0.
  \label{eq:time_evolution}
\end{equation}
Here $A_t = A(\Gamma_t) = A(\Gamma_t, 0) = A(\Gamma_0, t)$, where $\Gamma_0$ is an initial point in phase space and $\Gamma_t$ the state it is evolved into at time $t$.
The use of the imaginary $i$ is a widely used convention, originating from the quantum mechanics formalism, such that $\mathcal{L}$ is Hermite.
It is of no consequence here and one can consider the product $i \mathcal{L}$ as one operator.
The exact form of $i \mathcal{L}$ is of no importance for the general derivation.

In a classical mechanics setting the microscopic evolution can always be thought of as a trajectory through phase space, parameterized by $\Gamma_t$.
For convenience we introduce the following notation for the time-derivative
\begin{equation}
  \dot{A}_t = \frac{d}{dt} A(\Gamma_t) = \frac{d}{dt} A(\Gamma_t,0) = \frac{\partial}{\partial t} A(\Gamma_0, t) = i \mathcal{L} A_t.
\end{equation}
A useful operator identity given by Kawasaki \cite{Kaw73} is
\begin{multline}
  \frac{d}{dt} \exp[i \mathcal{L} \, t] = \exp[i \mathcal{L} \, t] \, i \mathcal{L}_0\\
  + \int_0^t ds \, \exp[i \mathcal{L} \, s] \, i \mathcal{L}_0 \, \exp[(i \mathcal{L}-i\mathcal{L}_0) \, (t-s) ] \, (i\mathcal{L}-i\mathcal{L}_0)\\ 
  + \exp[(i\mathcal{L}-i\mathcal{L}_0) \, t] \, (i\mathcal{L}-i\mathcal{L}_0).
  \label{eq:kawasaki_identity}
\end{multline}
The proof of this identity is straightforward when realizing that the integrand can also be written as
\begin{equation}
 \frac{d}{ds} \Bigl ( \exp[i\mathcal{L}\, s] \, \exp[(i\mathcal{L}-i\mathcal{L}_0) \, (t-s)] \, (i\mathcal{L}-i\mathcal{L}_0) \Bigr ).
\end{equation}
The identity is valid for any additional Liouville operator $\mathcal{L}_0$.

In projection operator formalism, $\mathcal{L}_0$ in Eq.~\eqref{eq:kawasaki_identity}, is the projected Liouville operator,
\begin{equation}
  i\mathcal{L}_0 = \mathcal{P} \, i\mathcal{L}.
  \label{eq:L0_def}
\end{equation}
Here $\mathcal{P}$ is the projection operator.
This operator will be specified further below.
For now we will just perform a formal exercise.

Using Eq.~\eqref{eq:time_evolution} for the macroscopic state $X_t=X(\Gamma_t)$, inserting this into Eq.~\eqref{eq:kawasaki_identity} and using definition Eq.~\eqref{eq:L0_def} gives,
\begin{multline}
  \dot{X}_t = \exp[i \mathcal{L} \, t] \, \mathcal{P} \, i \mathcal{L} \, X_0 + \\
  \int_0^t ds \, \exp[i \mathcal{L} \, s ] \, \mathcal{P} \, i\mathcal{L} \, \exp[(1-\mathcal{P}) \, i \mathcal{L} \, (t-s)] \, (1-\mathcal{P}) \, i \mathcal{L} \, X_0\\ + \exp[(1-\mathcal{P}) \, i\mathcal{L} t] \, (1-\mathcal{P}) \, i \mathcal{L} \, X_0.
  \label{eq:kawasaki_identity_applied}
\end{multline}
The last rhs term in Eq.~\eqref{eq:kawasaki_identity_applied}, is used as the definition of the fluctuating term
\begin{equation}
  f^{\mathrm{fluct}}(\Gamma_0, s) \equiv \exp[(1-\mathcal{P}) \, i \mathcal{L} s] \, (1-\mathcal{P}) \, i\mathcal{L} \, X_0.
  \label{eq:fluctuating_term}
\end{equation}
This term also arises inside the second integral term.
The value of $f^{\mathrm{fluct}}$ depends on the time $s$ passed since preparation in the initial state $\Gamma_0$.
The reason why Eq.~\eqref{eq:fluctuating_term} is referred to as the fluctuating term will become more apparent below.
Using definition Eq.~\eqref{eq:fluctuating_term}, one can rewrite Eq.~\eqref{eq:kawasaki_identity_applied} as,
\begin{multline}
  \dot{X}_t = \exp[i \mathcal{L} \, t] \, \mathcal{P} \, i \mathcal{L} \, X_0 + \\
  \int_0^t ds \, \exp[i \mathcal{L} \, s ] \, \mathcal{P} \, i\mathcal{L} \, f^{\mathrm{fluct}}(\Gamma_0,t-s)  + f^{\mathrm{fluct}}(\Gamma_0,t).
  \label{eq:decomposed_with_fluctuations}
\end{multline}

Note that up to this point we only performed a formal decomposition of the evolution equation and introduced some definitions.
No extra assumptions where introduced.

\subsection{The projection operator}

A convenient property of using Liouville operators is that one can consider ensembles of (initial) states.
An ensemble is characterized by a measure or density.
Measures (loosely speaking, statistical weights) are assigned to all microstates.
A quantity maps each microscopic state to a (finite vector of) real number(s).
The measure can be used to weigh these values.
Mathematically the measure characterizing the ensemble is the dual object of a quantity.
By using the measure one can map the values of a quantity corresponding to an ensemble of microstates to one (vector of) real number(s).
The full mathematical structure of ensembles and measures is called a sigma-algebra.

The space of possible measures can be restricted to represent probability measures.
During time evolution classical mechanics does not destroy the non-negative properties corresponding to probability measures.
The pairing of ensembles and quantities can therefore be interpreted as computing expectation values.
Note that these expectation values are not necessarily related to ``reality''.
Actually, ensembles and the corresponding measures are primarily a mathematical tool, because in reality a system is always in one microstate.

The mathematical structure of measures on ensembles can be used to decompose the equations of motion in a ``relevant'' and ``irrelevant'' part.
This is done in projection-operator formalism and will be outlined below.
The procedure is formal.
To derive at a useful decomposition, a restriction on the choice of relevant ensembles must be made.

In the derivation of linear generalized Langevin equations one usually starts by introducing a Hilbert space.
The projection operator is next defined using the inner product on this space.
For non-linear Langevin equations the introduction of a Hilbert space is not necessary.
So, we will not use this approach here.
This is the point where the current derivation starts to deviate from Kawasaki's derivation and becomes more similar to the early derivation of Zwanzig \cite{Zwa61}.
The difference with Zwanzig is that we will remain on the level of Langevin equations and will not attack the problem via the Fokker-Planck side.

A general (linear) projection operator of a microstate $\Gamma$ onto its corresponding macrostate $X(\Gamma)$ can be immediately introduced.
Operating on a quantity $A(\Gamma)$ the projection gives
\begin{equation}
  (\mathcal{P} A)(\Gamma)  = \int A(\Gamma') \, \mu[d\Gamma'| X(\Gamma') \in dX(\Gamma)].
  \label{eq:projection_operator}
\end{equation}
Here we used a measure theoretic notation of the integral.
The measure used is a conditional measure based on the, still to be defined, measure $\mu$.
With the notation, $X(\Gamma') \in dX(\Gamma)$ we mean the set of $\Gamma'$'s such that $X(\Gamma')$ is in a small, size $\epsilon$, measurable neighborhood around the macrostate $X(\Gamma)$.
We consider the limit $\epsilon \rightarrow 0$.
We assume that the conditional measure is well defined such that, in the limit $\epsilon \rightarrow 0$, the integrals using this conditional measure of a large class of sufficiently smooth functions exists (and the limit is smooth).

At this point one might suspect we are introducing the finite shell model that was argued against in the introduction.
There are two subtleties to be considered here.
Firstly, indeed, a $\epsilon$-region around macrostate $X$ is considered.
Depending on the choice of the measure in the microscopic space $\mu$ the corresponding region in microscopic phase space looks like a shell that has fixed with or is more 'wobbling'.
Note that at this point, however, the choice of $\mu$ is still completely open.
Secondly, the conditional measure remains well defined in the limit $\epsilon \rightarrow 0$.
Therefore the width is not finite.

The conditional measure is defined as 
\begin{equation}
  \mu[d\Gamma | X(\Gamma) \in dX] \equiv
  \frac{\mu[d\Gamma \cap \{ \Gamma: X(\Gamma) \in dX \}]}{\mu[\{\Gamma : X(\Gamma) \in dX \}]},
  \label{eq:conditional_measure}
\end{equation}
if $X(\Gamma) \in dX$, and zero otherwise.
It obeys, trivially, the property 
\begin{equation}
  \int \mu[d\Gamma'| X(\Gamma') \in dX(\Gamma)] = 1.
  \label{eq:conditional_measure_normalization}
\end{equation}
For Eq.~\eqref{eq:projection_operator} to define a projection operator it should leave properties that depend only on the macrostate, $A(X(\Gamma))$, unchanged.
Due to property Eq.~\eqref{eq:conditional_measure_normalization} this requirement is obeyed.
So, Eq.~\eqref{eq:projection_operator} is indeed a projection operation.
In principle any underlying measure, $\mu$, defines by means of Eq.~\eqref{eq:conditional_measure} a projection operator and by means of Eq.~\eqref{eq:kawasaki_identity_applied} a decomposition corresponding to this projection.

The conditional measure, Eq.~\eqref{eq:conditional_measure}, defines a generalized microcanonical ensemble.
All microstates, $\Gamma'$, consistent with a macrostate $X(\Gamma)$ contribute with a certain weight.
As alternative notation to Eq.~\eqref{eq:projection_operator} one can write it as a conditional expectation value,
\begin{equation}
  (\mathcal{P} A)(X) = E(A | X).
  \label{eq:expectation_value}
\end{equation}
Using the expectation value notation, gives, e.g., that
\begin{equation}
  \exp[i \mathcal{L} \, t] \, \mathcal{P} \, i \mathcal{L} \, X_0 = \exp[i \mathcal{L} \, t] \, E(\dot{X}|X_0) = E(\dot{X}|X_t).
  \label{eq:conditional_notation}
\end{equation}
This expression should be read as the conditional expectation value corresponding to macrostate $X_t$ of the instantaneous macroscopic phase space velocity, $\dot{X}=\dot{\Gamma}' \cdot \partial X(\Gamma')/\partial \Gamma'$ (where $X(\Gamma')=X_t$). 
Using this notation, Eq.~\eqref{eq:kawasaki_identity_applied}, can be rewritten as
\begin{equation}
  \dot{X}_t = E(\dot{X}|X_t) + \int_0^t ds \, E(\dot{f}^\mathrm{fluct}(\cdot,t-s)|X_s) + f^\mathrm{fluct}(\Gamma_0,t).
  \label{eq:Langevin_conditional_notation}
\end{equation}

It is worthwhile to spend some time on the interpretation of Eq.~\eqref{eq:Langevin_conditional_notation}.
By means of the definition of the fluctuating term, Eq.~\eqref{eq:fluctuating_term}, one knows that, when letting the projection operator $\mathcal{P}$ act on $f^{\mathrm{fluct}}$, it will give zero,
\begin{equation}
  E(f^{\mathrm{fluct}}(\cdot,t)|X(\Gamma)) = (\mathcal{P} f^{\mathrm{fluct}})(\Gamma,t) = 0.
  \label{eq:average_fluct}
\end{equation}
This explains the terminology ``fluctuation''.
Note that, to lose the fluctuation term in Eq.~\eqref{eq:Langevin_conditional_notation}, one needs to average over all possible initial microstates consistent with macrostate $X_0$.
Although a microstate is initially consistent with $X_0$ this does not mean that, at a later time, it is still consistent with $X_t=X(\Gamma_t)$.
The expectation values arising inside Eq.~\eqref{eq:Langevin_conditional_notation} are averages over microstates consistent with attained macrostate $X_t=X(\Gamma_t)$.
The averaging over initial conditions (consistent with $X_0$) therefore gives rise to an averaging over possible values $X_t$ evolved from initial microscopic states consistent with $X_0$ (but not attained in reality).

In Eq.~\eqref{eq:Langevin_conditional_notation} the initial time has a special significance.
Taking a different time-origin will give a different equation.
By taking the current time as the time-origin the time integral is zero.
So, for this special case one can write an alternative form of $\dot{X}_t$
\begin{equation}
  \dot{X}_t = E(\dot{X}|X_t) + f^\mathrm{fluct}(\Gamma_t,0).
  \label{eq:Langevin_conditional_notation_zero}
\end{equation}
This identity will also be used later on.

\subsection{The entropy definition}

Finite classical systems are defined on a phase space characterized by the coordinates and momenta of $n$ particles, so by $6n$ real numbers.
On a space $\mathcal{R}^{6n}$ one can define a Lebesgue measure.
The Lebesgue measure is based on the notion of a volume of hypercubes.
Using this basic definition it generalized this notion of volume to more elaborate sets, namely, members of the sigma-algebra.
For sufficiently smooth measures $\mu$ one can relate this measure to the Lebesgue measure as
\begin{equation}
  \mu(d\Gamma) = w(\Gamma) \, \mu_L(d\Gamma).
  \label{eq:lebesgue}
\end{equation}
For the proof given below the important property of the Lebesgue measure is that it is translational invariant.
If the smoothness conditions are not met, and the relevant measure might be a fractal one, then still a splitting as given by Eq.~\eqref{eq:lebesgue} might be possible with a translation invariant fractal measure instead of the Lebesgue measure.

To relate the conditional measure to the Lebesgue measure also the denominator in Eq.~\eqref{eq:conditional_measure} should be related to a Lebesgue measure.
This is the point at which the entropy is introduced.
Here,
\begin{multline}
  \mu[\{\Gamma:X(\Gamma') \in dX \}] = \\ \int \mu[d\Gamma \cap \{ \Gamma: X(\Gamma) \in dX \}] \equiv \exp[ S(X) ] \, \mu_L(dX),
  \label{eq:entropy_definition2}
\end{multline}
if $X(\Gamma) \in dX$ and zero otherwise.
Note that the Lebesgue measure that appears is defined on the macroscopic space as opposed to the microscopic one.
It arises because,  if $\mu$ is sufficiently smooth, then the sum of measure of two small neighboring (non-overlapping) set equals the total measure of the union of these sets.
All possible pre-factors are incorporated in the entropy definition.
One can read Eq.~\eqref{eq:entropy_definition2} as follows.
The factor $\exp[S(X)]$ is the measure of microscopic phase space per unit volume (Lebesgue measure) macroscopic phase space.
Note that if $\mu$ was not sufficiently smooth one still might have been able to introduce an entropy if one could find a suitable translational invariant fractal measure on the macroscopic phase space.

Using the definitions Eq.~\eqref{eq:lebesgue} and Eq.~\eqref{eq:entropy_definition2} and the formal definition of the Dirac delta function one can write the conditional measure as
\begin{multline}
  \mu[d\Gamma| X(\Gamma) \in dX] = \\
  w(\Gamma) \, \exp[-S(X)] \, \delta[X(\Gamma)-X] \, \mu_L(d\Gamma).
  \label{eq:conditional_measure_2}
\end{multline}
This expression is convenient because it clearly separates the parts that depend on $\Gamma$ directly and others that only depend on $\Gamma$ through the macrostate, i.e., via $X(\Gamma)$.

In Eq.~\eqref{eq:Langevin_conditional_notation} conditional expectation values of the form $E(\dot{A}|X_t)$ play an important role.
Using the form given by Eq.~\eqref{eq:conditional_measure_2} combined with Eq.~\eqref{eq:expectation_value} and Eq.~\eqref{eq:projection_operator} a second representation can be found as
\begin{widetext}
\begin{equation}
  \begin{split}
    E(\dot{A}|X) &= \, \exp[-S(X)] \, \int \mu_L(d\Gamma) \, w(\Gamma) \, \dot{\Gamma} \cdot \frac{\partial A}{\partial \Gamma } \, \delta[X(\Gamma) - X]\\ 
    &= - \, \exp[-S(X)] \, \int \mu_L(d\Gamma) \, A(\Gamma) \, \frac{\partial}{\partial \Gamma} \cdot \bigl ( \dot{\Gamma} w(\Gamma) \bigr ) \, \delta[X(\Gamma) - X] \\
    & \quad +  \, \exp[-S(X)] \, \frac{\partial}{\partial X} \cdot \int \mu_L(d\Gamma) \, \dot{X} (\Gamma) \, A(\Gamma) \, w(\Gamma) \, \delta[X(\Gamma) - X]\\
    &= - E \Bigl ( w^{-1} f \, \frac{\partial}{\partial \Gamma} \cdot \bigl ( \dot{\Gamma} w \bigr ) \Bigr | X \Bigr ) + \, \exp[-S(X)] \, \frac{\partial}{\partial X} \cdot \Bigl( \exp[S(X)] \, E(\dot{X} \, A | X) \Bigr).
  \end{split}
  \label{eq:alternative_dotA}
\end{equation}
\end{widetext}

Now comes the main mathematical point of this paper.
The final expression of Eq.~\eqref{eq:alternative_dotA} consists of two terms.
In most historic derivations the first term is taken to be zero.
There are good reasons for doing this, but it is usually done without explicit mentioning it.
What we want to do here is to discuss when this term is zero, and what are the consequences for the entropy expression.
Clearly the term is zero when $\mu$ is an invariant measure, i.e., if
\begin{equation}
  \frac{\partial}{\partial \Gamma} \cdot [\dot{\Gamma} w(\Gamma)] = 0.
  \label{eq:invariant_measure}
\end{equation}
Depending on the ergodic properties of the dynamics there are one or many invariant measures.
Independent of the detailed dynamics, in classical mechanics,
\begin{equation}
  w(\Gamma) = 1,
  \label{eq:w_is_one}
\end{equation}
is always a valid choice.

The reason is the Liouville theorem.
For a closed classical system described by Hamilton equations the Liouville theorem, i.e. incompressibility of phase space, holds.
In mathematical terms this is expressed as $\partial/\partial \Gamma \cdot \dot{\Gamma} =0$.
So, Eq.~\eqref{eq:invariant_measure} is always obeyed for constant $w$.
This choice determines, by means of Eq.~\eqref{eq:conditional_measure_2}, the conditional measure.
And consequently, also the entropy, by Eq.~\eqref{eq:entropy_definition2}.

Taking $w$ to be invariant, the first term in Eq.~\eqref{eq:alternative_dotA} cancels, and Eq.~\eqref{eq:Langevin_conditional_notation} becomes
\begin{multline}
  \dot{X}_t = E(\dot{X}|X_t) + \int_0^t ds \, \exp[-S(X_s)]\\ \times \frac{\partial}{\partial X_s} \cdot \Bigl( \exp[S(X_s)] \,
  E(\dot{X} \, f^\mathrm{fluct}(\cdot,t-s) | X_s) \Bigr)\\ + f^\mathrm{fluct}(\Gamma_0,t).
  \label{eq:Langevin_conditional_notation_invariant}
\end{multline}

Combining the properties Eq.~\eqref{eq:average_fluct} and Eq.~\eqref{eq:Langevin_conditional_notation_zero} one finds that
\begin{multline}
  E(\dot{X} \, f^\mathrm{fluct}(\cdot,t-s) | X_s) = \\ E( [E(\dot{X}|X_s) + f^\mathrm{fluct}(\cdot,0)] \, f^\mathrm{fluct}(\cdot,t-s) | X_s) \\ = E( f^\mathrm{fluct}(\cdot,0) \, f^\mathrm{fluct}(\cdot,t-s) | X_s),
\end{multline}
so
\begin{multline}
  \dot{X}_t = E(\dot{X}|X_t) + \int_0^t ds \, \exp[-S(X_s)]\\ \times \frac{\partial}{\partial X_s} \cdot \Bigl( \exp[S(X_s)] \,
   E( f^\mathrm{fluct}(\cdot,0) \, f^\mathrm{fluct}(\cdot,t-s) | X_s) \Bigr)\\ + f^\mathrm{fluct}(\Gamma_0,t).
  \label{eq:generalized_Langevin}
\end{multline}
This is the generalized nonlinear Langevin equation.
The shape of the equation clearly illustrates the fluctuation-dissipation relation.
This equation is generally valid.
The only ingredient in this derivation, besides straightforward formal mathematical manipulation, is that for the entropy definition an invariant measure was used.

The derivation is valid for general closed systems.
Note that open systems assume that it is possible to make a division between system and environment.
This separation is always an approximation.
In the current setting we will (realistically) describe the total system plus environment as a closed system.
The full microscopic description of the whole is assumed to obey Liouville's theorem.
Therefore, Eq.~\eqref{eq:generalized_Langevin} is always valid for the whole.
In the macroscopic description the environment might be modeled in a very elementary fashion, e.g., as a heat-bath.
The entropy that appears in Eq.~\eqref{eq:generalized_Langevin} is, in the case of an open system, the total entropy of the system plus environment.
This means that what is called entropy in the current paper is, for different kinds of environments a quantity proportional to, what is usually referred to as free-energy or Gibbs energy, available energy or exergy etc..

\section{The generalized microcanonical ensemble and entropy}

The ensemble given by Eq.~\eqref{eq:conditional_measure_2}, with $w(\Gamma)=1$, is a generalized microcanonical distribution corresponding to macroscopic state $X$.
This is a straightforward generalization of the energy based microcanonical ensemble.
There are, however, some features that are worthwhile pointing out.

Firstly, by means of Liouville's theorem and to obtain a useful generalized nonlinear Langevin equation, $w(\Gamma)=\text{constant}$, is the only sensible choice one can make without any extra knowledge on the dynamics.

The energy based ensemble is usually defined by introducing a finite, but small, shell thickness $\epsilon_0$.
Because of conservation of energy the system remains within the shell.
Traditionally the Liouville theorem,  combined with a reasoning on ergodicity, is used to motivate the microcanonical ensemble.
It might be worthwhile to note that the microcanonical ensemble is often used to determine thermodynamic behavior when changing the energy.
So, the energy is not constant!
It is a (slow) dynamic variable.
If one considers a total energy that can change, it is difficult to motivate why the same thickness $\epsilon_0$ should be the same for the shells at different $E$.
These kind of conceptual problems are not present in the current derivation.
The choice for the entropy definition is such that a very inconvenient term in the generalized Langevin equation cancels.
No ergodic reasoning is used.
What remains to be shown is that the generalized non-linear Langevin equation is useful.

In the current case $X$ is a dynamic variable.
There is no a priori assumption about its nature.
In the definition of the generalized microcanonical distribution, Eq.~\eqref{eq:conditional_measure_2}, there seems to be a preferred coordinate system (of the macroscopic state) introduced because of the definition of the delta-function.
Because of a kind of the choice $w=1$ the delta-function defines a (infinitesimal) shell with unit thickness (per unit $dX$) in the microscopic phase space.
One might wonder what the behavior upon coordinate transformation $X \rightarrow Y$ is.
The reason one might, at that point, get confused when interpreting, Eq.~\eqref{eq:conditional_measure_2}, is because one thinks of the entropy as a scalar quantity.
As can be seen from Eq.~\eqref{eq:entropy_definition2}, if one considers a smooth one-on-one transformation, $X \rightarrow Y(X)$, of the macroscopic space then,
\begin{multline}
  \exp[S(X)] \, \mu_L(dX) = \mu[\{\Gamma:X(\Gamma) \in dX \}]\\
  = \mu[\{\Gamma:Y(X(\Gamma)) \in \frac{\partial Y}{\partial X} \cdot dX \}] \\
  = \exp[ S(Y(X)) ] \, \left| \det \biggl (\frac{\partial Y}{\partial X} \biggr ) \right | \, \mu_L(dX),
\end{multline}
or
\begin{equation}
  S(Y) = S(X) - \ln \left | \det \biggl (\frac{\partial Y}{\partial X} \biggr ) \right |.
  \label{eq:coordinate_transform}
\end{equation}
This illustrates that upon a change of variables the entropy does not transform as a variable.

One can make the choice not to accept this naturally arising, non-scalar, scaling behavior.
If one wants, for some reason, to consider entropy as a scalar quantity then there is a preferred ``coordinate'' system where Eq.~\eqref{eq:generalized_Langevin} holds with entropy $S(X)$.
If one now considers another parameterization and takes, entropy to be scalar, $S(Y)=S(X)$, then Eq.~\eqref{eq:generalized_Langevin} has extra determinant terms due to the coordinate transformation.
If one does not want these terms to arise then Eq.~\eqref{eq:coordinate_transform} should be used.

The traditional interpretation of $\exp[S(X)]$ is the number of microscopic states for a macroscopic state $X$.
The interpretation in this paper is to view $\exp[S(X)]$ as the Liouville measure of microscopic space per unit (Lebesgue) volume macroscopic space.
This is more easily defined (at least in classical theory) because states can not be counted, but volume is well defined.
The mathematics shows that this interpretation also gives the simplest equations.
The form of the equation for the generalized Langevin equation, Eq.~\eqref{eq:generalized_Langevin}, is independent of the chosen parameterization.

\section{Stochastic differential equations}

The generalized Langevin equation, Eq.~\eqref{eq:generalized_Langevin}, is a formal decomposition of the microscopic equations of motion.
It contains no new information.
Full expressions of the fluctuating term $f^\mathrm{fluct}(\Gamma_0,t)$ are very complicated.
Its use lies in the fact that it can be used as a starting point for approximations.

Suitable choices for the macroscopic variables $X$ can be made.
The usual approach is to choose the variables such the remainder characterized by $f^\mathrm{fluct}(\Gamma_0,t)$ decorrelates quickly.
If this is the case the most simple modeling assumption for the fluctuation term $f^\mathrm{fluct}(\Gamma_0,t)$ is that it is white noise, i.e., a stochastic Gaussian process with decorrelation time 0.

In reality, of course, there is a finite decorrelation time $\tau$.
The modeling assumption is that (complete) decorrelation is very fast, i.e., the change of $X$ is very small on the time scale $\tau$.
One is interested in phenomena on time scales much bigger than $\tau$.
For time scales $\Delta t \gg \tau$ the $f^\mathrm{fluct}$ can be modeled by means of a Wiener process $W$,
\begin{equation}
  \int_0^{\Delta t} f^\mathrm{fluct}(\Gamma_0,t) \, dt \approx \sqrt{2 \mathbf{D}} \cdot \Delta W,
  \label{eq:white_noise}
\end{equation}
where $\mathbf{D}$ is a positive definite matrix.
A Wiener process is a Gaussian stochastic process.
Each increment over a time-step $\Delta t$ has zero average and variance $\Delta t$,
\begin{equation}
  \langle \Delta W \rangle = 0, \text{ and } \langle \Delta W \otimes \Delta W \rangle = \mathbf{I} \Delta t.
\end{equation}
Increments over non-overlapping time intervals are statistically independent.
The stochastic term on the rhs of Eq.~\eqref{eq:white_noise} should be read using the so-called Ito-interpretation (see, e.g., \cite{Ott96}).
This means that the expectation value of the increment, averaged over initial conditions of $\Gamma_0$ is zero.
This is a consequence of Eq.~\eqref{eq:average_fluct}.

When integrating Eq.~\eqref{eq:generalized_Langevin} for $\Delta t$ one finds
\begin{multline}
  \Delta X  = \int_0^{\Delta t} dt \, E(\dot{X}|X_t) + \\
  \int_0^{\Delta t} ds \, \exp[-S(X_s)] \, \frac{\partial}{\partial X_s} \cdot \Bigl( \exp[S(X_s)] \\ 
  \times \int_{0}^{\Delta t-s} \! \! ds' \, E( f^\mathrm{fluct}(\cdot,0) \, f^\mathrm{fluct}(\cdot,s') | X_s) \Bigr)\\
  +  \int_0^{\Delta t} dt f^\mathrm{fluct}(\Gamma_0,t).
  \label{eq:Langevin_integrated}
\end{multline}
Since the fluctuating force is decorrelating quickly one finds that the integral
\begin{equation}
  \tilde{\ten{D}}(X_s) = \int_{0}^{\Delta t-s} \! \! ds' \, E( f^\mathrm{fluct}(\cdot,0) \, f^\mathrm{fluct}(\cdot,s') | X_s),
  \label{eq:dissipation}
\end{equation}
does not depend on $s$, other than through $X_s$, except when $s=\Delta t-\mathcal{O}(\tau)$.
The diffusion coefficient as defined by Eq.~\eqref{eq:white_noise} can be computed from the variance of the fluctuating term.
This gives
\begin{multline}
  \ten{D}(X_0) = \frac{1}{2 \Delta t} \int_{0}^{\Delta t} \int_{0}^{\Delta t}  \! \! ds \, ds' \\
  \times E( f^\mathrm{fluct}(\cdot,s) \, f^\mathrm{fluct}(\cdot,s') | X_0).
  \label{eq:variance_fluctuation}
\end{multline}
Comparing Eq.~\eqref{eq:dissipation} and Eq.~\eqref{eq:variance_fluctuation} one sees that both quantities are similar, but not exactly the same.

To be able to establish, rigorously, the usual fluctuation-dissipation relation one needs to introduce the extra assumption that $X_t$ is a slow variable.
During a few decorrelation times $\tau$, $X_t$ has hardly changed.
This assumption gives that
\begin{multline}
  E( f^\mathrm{fluct}(\cdot,0) \, f^\mathrm{fluct}(\cdot,s') | X_s)\\ \approx E( f^\mathrm{fluct}(\cdot,0) \, f^\mathrm{fluct}(\cdot,s') | X_0)\\ \approx E( f^\mathrm{fluct}(\cdot,s'') \, f^\mathrm{fluct}(\cdot,s'+s'') | X_s). 
  \label{eq:time-translation-invariance}
\end{multline}
for $s, \, s', s'' = O(\Delta t)$.
Using this time translation invariance, and computing the expectation value of the variance of Eq.~\eqref{eq:white_noise}, gives
\begin{equation}
  2 \mathbf{D}(X_0) = \tilde{\mathbf{D}}(X_0) + \tilde{\mathbf{D}}^T(X_0).
  \label{eq:symmetric}
\end{equation}
So, if $\tilde{\mathbf{D}}$ is symmetric both are equal.
In the general case, however, one can write
\begin{equation}
  \tilde{\mathbf{D}} = \mathbf{D} + \mathbf{A},
\end{equation}
where $\mathbf{A}$ is anti-symmetric, i.e., $\mathbf{A}^T = -\mathbf{A}$.

If one imagines $X$ as a vector of values then $\tilde{\mathbf{D}}$ is a matrix.
For suitable chosen macroscopic variables individual elements on both sides of its diagonal are, either, symmetric ($D_{ij}=D_{ji}$) or anti-symmetric ($D_{ij}=-D_{ji}$).
Suitably means that a time-reversal operation is well defined on both the microscopic and macroscopic space.
If $\Gamma^*$ is the time reversed microscopic state corresponding to $\Gamma$, then the operation $X(\Gamma^*) = X^*(\Gamma)$ should make sense.
Assuming the time-translation invariance as given in Eq.~\eqref{eq:time-translation-invariance} one finds that
\begin{multline}
  E( f^\mathrm{fluct}(\cdot,0) \, f^\mathrm{fluct}(\cdot,\Delta s) | X)\\
  \begin{aligned}
    &= E( f^\mathrm{fluct}(\cdot, -\Delta s) \, f^\mathrm{fluct}(\cdot, 0) | X)\\
    &= E( f^{\mathrm{fluct}}(\cdot, \Delta s) \, f^{\mathrm{fluct}}(\cdot, 0) | X^*)\\
    &= E( f^{\mathrm{fluct}}(\cdot, 0) \, f^{\mathrm{fluct}}(\cdot,\Delta s) | X^*)^T.  
  \end{aligned}
\end{multline}
Depending on the parity of the time-reversal of individual components of the fluctuating contributions, i.e., $f^{\mathrm{fluct}}_i(X^*) = \pm f^{\mathrm{fluct}}_i(X)$, cross correlations of terms with opposite parities contribute to the anti-symmetric matrix $\mathbf{A}$, all others to $\mathbf{D}$. 

These anti-symmetric contributions correspond to the extension of Onsager's reciprocal principle by Casimir \cite{Cas45}.
Clearly, $\tilde{\mathbf{D}}$ can always be decomposed in a symmetric and antisymmetric part.
From Eq.~\eqref{eq:symmetric} follows that the symmetric part is directly related to the fluctuations in the stochastic differential equation.
The Casimir relations can be used to that the antisymmetric part is zero (or at least to determine its rank).
If $\mathbf{A}$ is nonzero, it makes sense to choose a preferred macroscopic coordinate system such that $\mathbf{A}$ only acts on a small subspace of the vector space.

Under the assumption of rapid decorrelating fluctuations, the generalized Langevin equation, Eq.~\eqref{eq:generalized_Langevin}, can be simplified to a stochastic differential equation.
First one considers, Eq.~\eqref{eq:generalized_Langevin}, for $\Delta t \gg \tau$.
Then use the modeling assumption that $X_t$ is slow, that macroscopic variables are chosen such that one can perform time-reversal and that the fluctuating term can be modeled as Gaussian noise.
One obtains a stochastic difference equation (strictly valid after integration over $\Delta t \gg \tau$), which can be well approximated by the stochastic differential equation
\begin{equation}
  \begin{split}
  dX_t & = E(\dot{X} | X_t) \, dt + \exp[-S] \\
  & \times \frac{\partial}{\partial X_{t}} \cdot \biggl [ \tilde{\mathbf{D}} \, \exp[S] \biggr ] \, dt + \sqrt{2 \mathbf{D}} \cdot dW_t \\ 
  & = E(\dot{X} | X_t) \, dt + \tilde{\mathbf{D}}^T \cdot \frac{\partial S}{\partial X_{t}} \, dt + \frac{\partial}{\partial X_{t}} \cdot \tilde{\mathbf{D}} \, dt \\
  &\quad  + \sqrt{2 \mathbf{D}} \cdot dW_t.
  \end{split}
  \label{eq:SDE}
\end{equation}
This stochastic differential equation has three main contributions a instantaneous (reversible) part, an irreversible (dissipative) contribution and a fluctuating (random) part.
The first term on the rhs gives the instantaneous change of $X_t$ averaged over all possible microstates consistent with this state.
The last term models the fluctuations with respect to this average motion.
On time scales larger than decorrelation time, $\tau$, this is effectively modeled by means of a white noise, or Wiener, process.
The irreversible term gives a drift toward macrostates with higher entropy.
This bias can be explained intuitively by the argument that these regions correspond to a larger micro-phase-space-volume.
Therefore the ``residence time'' is longer.
 
\section{GENERIC equations}

This Langevin equation, Eq.~\eqref{eq:SDE}, is very similar to the governing equation of the GENERIC formalism.
In fact it is the same except the GENERIC formalism imposes extra structure on the matrices (or more general operators) arising in the formula.
The GENERIC equation has the form, \cite{Ott05},
\begin{equation}
  dX_t = \mathbf{L} \cdot \frac{\partial H}{\partial X_{t}} \, dt + \mathbf{D} \cdot \frac{\partial S}{\partial X_{t}} \, dt + \frac{\partial}{\partial X_{t}} \cdot \mathbf{D} \, dt + \sqrt{2 \mathbf{D}} \cdot dW_t.
  \label{eq:GENERIC-equation}
\end{equation}
The most strict assumption of the GENERIC formalism is the structure of the reversible part.
In the terminology of the formalism it is a two generator equation where the generators are energy $H$ and entropy $S$.
The motivation to introduce the ``Poisson matrix'' $L$ and the energy $H$ is that this gives a nice structure to the equations.
In fact, within the GENERIC formalism the reversible part obeys Hamilton dynamics, or more general obeys the underlying geometric structure, the Poisson structure.
Within the framework two additional degeneracy conditions, that will be discussed below, are also obeyed.

If we write the microscopic dynamics as
\begin{equation}
  \dot{\Gamma} = \mathbf{L}^\mathrm{micro} \cdot \frac{\partial H^\mathrm{micro} (\Gamma)}{\partial \Gamma}.
\end{equation}
The Poisson structure can be expressed in terms of properties of the Poisson matrix as
\begin{equation}
   \mathbf{L}^\mathrm{micro} =  -(\mathbf{L}^\mathrm{micro})^T \text{ and } \mathbf{L}^\mathrm{micro} \cdot \frac{\partial \mathbf{L}^\mathrm{micro}}{\partial \Gamma} = 0.
   \label{eq:Poisson-structure}
\end{equation}

The GENERIC equations can be found if one assumes that Hamiltonian of the system can be expressed in terms of macroscopic variables,
\begin{equation}
  H^\mathrm{micr}(\Gamma) = H(X(\Gamma)).
  \label{eq:macroscopic_energy}
\end{equation}
In that case,
\begin{equation}
  \begin{split}
    E(\dot{X} | X_t)  &= E \biggl ( \frac{\partial X}{\partial \Gamma} \cdot \mathbf{L}^\mathrm{micro} \cdot \frac{\partial H^\mathrm{micro} (\Gamma)}{\partial \Gamma} \biggl | X_t  \biggr ) \\
    &= E \biggl ( \frac{\partial X}{\partial \Gamma} \cdot \mathbf{L}^\mathrm{micro} \cdot \frac{\partial X}{\partial \Gamma} \biggl | X_t  \biggr ) \cdot \frac{\partial H}{\partial X_t}\\
    &= \mathbf{L} \cdot \frac{\partial H}{\partial X_t}.
      \label{eq:GENERIC_approx}
  \end{split}
\end{equation}
It is clear that the coarse-grained Poisson matrix is anti-symmetric.
In many theories the form Eq.~\eqref{eq:macroscopic_energy} is put in by design.
Usually the energy can be approximated by using quantities such as the kinetic energy of the center-of-mass of a group of atoms etc..
The remaining energy not covered by these contributions are collected into a new variable, namely, the internal energy.

Not all equations are easily put into the GENERIC form.
The most elementary example is the Brownian motion of a particle in a background velocity field.
If the motion is described by considering position only, i.e., the momentum variable is eliminated, then
\begin{equation}
  d \vec{X} = \vec{v}(\vec{X}) \, dt + \sqrt{2 D} \, d\vec{W}.
\end{equation}
It is hard, when not allowed to use a momentum of the particle as a variable, to write down its energy.
It is clear, however, that the expectation value of the instantaneous velocity of a Brownian particle is the background fluid velocity, $E(\dot{\vec{X}}|\vec{X})=\vec{v}(\vec{X})$.

Let's assume that an macroscopic Hamiltonian can be defined, e.g., by introducing an internal energy.
The second property in Eq.~\eqref{eq:Poisson-structure}, that is equivalent to the Jacobi property of Poisson brackets, i.e., 
\begin{equation}
  \mathbf{L} \cdot \frac{\partial \mathbf{L}}{\partial X} = 0,
  \label{eq:Jacobi-identity}
\end{equation}
is less trivial to prove in the coarse-grained case.
It is acknowledged by the people involved in the GENERIC movement that the Jacobi identity can not be proved in general (yet) \cite[p. 235]{Ott05}.
Clearly, the Jacobi-identity is obeyed when $\mathbf{L}$ is independent of the macroscopic state $X$.
This is almost always the case for the classical macroscopic transport equations such as the Navier-Stokes equation.

The GENERIC-movement has, however, started a certification program, \cite{Ott05}, to check, among other properties, whether macroscopic theories have the full Poisson structure.
If a macroscopic equation does not have it, they claim it is thermodynamically inconsistent.
This severe judgment seems not fully justified since the Jacobi-identity is not proved.
An example, where a well established equation of motion was supposedly disproved, is the case of the Doi-Edwards reptation model without independent alignment \cite{Ott99}.
This conclusion was disputed in \cite{Mil04} and defended in \cite{Edw04}.
It is hard to see who is right and why.
We conclude, a bit provocative, with the conclusion that for most coarse-graining purposes the rigorous result $E(\dot{X}|X)$ seems sufficient and the GENERIC expression, with its assumptions that need not be obeyed, seems to mainly add confusion.

\subsection{Degeneracy conditions}

Besides the Poisson structure the GENERIC formalism also prescribes two, so called, degeneracy conditions.
Before discussing these we will give a degeneracy condition that follows directly from the projection operator formalism.
From the current operator formalism imposes a restriction on the instantaneous or, reversible, contribution.
Applying Eq.~\eqref{eq:alternative_dotA} for $A=1$ (and $w$ invariant) gives,
\begin{equation}
  \exp[-S] \frac{\partial}{\partial X_t} \cdot \biggl [ E (\dot{X} | X_t) \, \exp[S] \biggr ] = 0.
  \label{eq:divergence_free}
\end{equation}
This is the consequence of Liouville's theorem (or more generally the existence of an invariant measure) applied to the macroscopic space.
One might interpret it as the fact that $E(\dot{X}| X_t)$ is ``divergence free''.
Here the volume-form defined by the entropy can be used in the divergence definition.
This is similar to the occurrence of a term $\sqrt{\det g}$, where $g$ is a metric, in Riemannian geometry definition of the divergence operator.

This condition can be rewritten as,
\begin{equation}
  E(\dot{X}| X_t ) \cdot \frac{\partial S}{\partial X_t} = - \frac{\partial}{\partial X_t} \cdot E(\dot{X}| X_t).
  \label{eq:entropy_production}
\end{equation}
Within the GENERIC formalism, Eq.~\eqref{eq:GENERIC-equation}, this equals,
\begin{equation}
  \left ( \mathbf{L} \cdot \frac{\partial H}{\partial X_t}\right )  \cdot \frac{\partial S}{\partial X_t} = - \frac{\partial}{\partial X_t} \cdot \left ( \mathbf{L} \cdot \frac{\partial H}{\partial X_t} \right ).
\end{equation}
By means of the anti-symmetry of matrix $\mathbf{L}$ one finds that
\begin{equation}
  \left ( \mathbf{L} \cdot \frac{\partial S}{\partial X_t} - \frac{\partial}{\partial X_t} \cdot \mathbf{L} \right ) \cdot \frac{\partial H}{\partial X_t} = 0.
  \label{eq:degeneracy_L}
\end{equation}
Since, within the GENERIC formalism, the expression for $\mathbf{L}$ is independent of $H$ the bracketed expression itself equals zero.
In the original papers on the GENERIC formalism, \cite{Grm97,Ott97_2}, the second term in Eq.~\eqref{eq:degeneracy_L} was not included.
In the book \cite[p. 233]{Ott05} it is found to be present.

Another instance where Eq.~\eqref{eq:entropy_production} arises is in (computational) studies where reversible thermostats are used, \cite{Ev02}.
In this case the evolution equation of a microscopic system is extended by one dissipative term that makes sure that the total kinetic energy (iso-kinetic thermostat) or the total energy (ergostat) stays constant.
This dissipative term causes Liouville's theorem not to be obeyed.
Usually arguments connected to the work done upon the system related to thermodynamic expression are used to derive the entropy production as given by, Eq.~\eqref{eq:entropy_production}.

The point of view in the current paper is that the full system, i.e., the microscopic thermostated system plus the environment (thermostat and driving force) obeys microscopic dynamics in reality.
The thermostat and the out-of-equilibrium driving force are a model of the environment.
This modeled system should be consistent with the underlying microscopic dynamics.
Since no fluctuating forces are incorporated in the reversible thermostat model this consistency condition leads to Eq.~\eqref{eq:entropy_production}. 

Note that this point of view also holds for the possible anti-symmetric part of $\tilde{\mathbf{D}}$ in Eq.~\eqref{eq:SDE}.
When applying Eq.~\eqref{eq:divergence_free} to the irreversible term in form of the first line in Eq.~\eqref{eq:divergence_free}, one finds that it is trivially obeyed also for this term because
\begin{equation}
  \frac{\partial^2}{\partial X \partial X} : ( \mathbf{A} \exp[S] ) = 0,
\end{equation}
due to the anti-symmetry of $\mathbf{A}$.
Note that on the level of a macroscopic equation, by considering this degeneracy condition only, it is hard to distinguish between the contribution of the instantaneous change, i.e., $E(\dot{X}|X)$, and the contribution of the anti-symmetric part of the entropy driven term.
In ``GENERIC-speak'' this latter term is driven by the entropy as generator and not by the energy.
The possibility of a non-zero anti-symmetric part driven by entropy seems to be missing from the GENERIC framework.

The second degeneracy condition in the GENERIC formalism is
\begin{equation}
  \mathbf{D} \cdot \frac{\partial H}{\partial X} = 0.
  \label{eq:degeneracy_D}
\end{equation}
This equality ensures that the time derivative of the total energy is zero.
The requirement is, however, stronger than necessarily needed for this purpose.
If, Eq.~\eqref{eq:macroscopic_energy} holds then, because of conservation of energy,
\begin{equation}
  0 = i \mathcal{L} H = (i \mathcal{L} X) \cdot \frac{\partial H}{\partial X},
\end{equation}
By means of the definition of the fluctuating force, Eq.~\eqref{eq:fluctuating_term}, and the definition of $\mathbf{D}$, Eq.~\eqref{eq:variance_fluctuation}, the degeneracy condition Eq.~\eqref{eq:degeneracy_D} can be proved.

\section{Non-equilibrium thermodynamics}

The goal of non-equilibrium thermodynamics is to supply a description of the time-evolution of a system in terms of coarse-grained, meso- or macroscopic, variables.
The generalized non-linear Langevin equation, after approximation for the fluctuating forces, supplies such a description.

In case of the derivation of the stochastic differential equation, Eq.~\eqref{eq:SDE}, the approximations are simple.
Fluctuations are micro-reversible and they decorrelate quickly.
The motion in microscopic phase space is reversible.
Because, in macroscopic space the volume of microscopic phase space corresponding to a unit volume macroscopic space needs not be constant the result is a bias.
The only reason is the mapping.
This bias is toward macroscopic states corresponding to more micro phase space volume (per unit macro phase space volume). 
This is exactly into accordance with the ordinary reasoning why systems tend toward increasing entropy.
The generalized nonlinear Langevin equation quantifies this tendency.
Therefore equilibrium thermodynamics results from it.

\subsection{Ergodicity and decorrelation}

Ergodicity is no requirement for Eq.~\eqref{eq:generalized_Langevin} to be valid.
The generalized Langevin equation is formally equivalent to the microscopic dynamics.
Ergodicity arguments come into play if one wants to approximate Eq.~\eqref{eq:generalized_Langevin}.
For example, if one wants to model the fluctuating forces by means of stochastic processes.

The property that there is one unique invariant probability measure, defined via Eq.~\eqref{eq:invariant_measure}, consistent with the dynamics of a system is called ergodicity (roughly speaking).
Let's call the probability distribution that defines this measure $\mu_\mathrm{erg}$.
Under, certain mild assumptions on the quantity $A(\Gamma)$, the Birkhoff ergodic theorem, can be derived 
\begin{equation}
  \lim _{T \rightarrow \infty} \frac{1}{T} \int_{0 \rightarrow T} A(\Gamma_t) \, dt = \int A(\Gamma) \, \mu_\mathrm{erg}(d\Gamma).
  \label{eq:ergodicity}
\end{equation}
This means long time averages are always equal to ensemble averages using $\rho_\mathrm{erg}$.
If a measure is unique, a single trajectory connects all points in phase space (except, possibly, for a set with measure 0).
Eq.~\eqref{eq:invariant_measure} transports the $w(\Gamma_0)$ defined at an initial point $\Gamma_0$ to all other points in phase space.

In the case of classical mechanics there is always one invariant measure, namely the Liouville measure.
This is usually not a probability measure since it can not be normalized.
Because of the conservation of energy (and other quantities) microscopic trajectories are restricted to constant (total) energy surfaces.
Conservation of energy causes that space can be decomposed in (dynamically) not-connected shells.
In this sense the Liouville measure is not unique.

Note that this non-uniqueness of the invariant measure should not be taken too serious.
As remarked earlier, also for the microcanonical system one is often interested in the change of the system when the energy is changes.
In this case the energy is not fixed, but a very slow variable, so the shells are, in fact, connected.

Usually, in classical mechanics, when discussing ergodicity one considers the ergodicity properties on a (total) energy shell.
This energy shell is then divided into a system and a heatbath.
The heatbath is usually taken to be very large.
Due to its largeness most of the (accessible) microscopic phase space corresponding to it can not be sampled within a finite time.
Moreover, it is clearly not realistic to model such a large system, say a laboratory with people walking around, as described by a microcanonical ensemble.

Note that the validity of Eq.~\eqref{eq:generalized_Langevin} does not depend on ergodicity, but its usefulness might in some way depend on it.
Non-uniqueness of the invariant measure is due to the possibility of decomposing microscopic phase space into invariant subspaces.
A trajectory starting in such a subspace will always remain in it.
If the variables $X$ are chosen such that they can not parameterize the invariant subspaces there might be a lasting dependence of the initial microscopic state a system starts in.

For example, for the same macrostate $X$ the microstate $\Gamma$ might be in subspace $A$ or $B$ and will stay there indefinitely.
In the two subspaces the decorrelation might occur differently.
If this is the case it will show up in the dynamics of $X$ but can not be modeled on the level of $X$.
In this case certain components of $f_t^\mathrm{fluct}$ do not decorrelate at all.
There is a lasting dependence on the initial microstate.

This non-ergodic behavior is the most extreme case.
In a dynamical situation, depending on which scale one is looking, there is little difference in not decorrelating or decorrelating very slowly.
In a dynamic theory (local) equilibrium is only an approximation.
What is important, for the usefulness of Eq.~\eqref{eq:generalized_Langevin} in devising approximate equations, is the decorrelation behavior of the fluctuation terms $f_t^\mathrm{fluct}$.
The ideal choice of variables $X$ is such that decorrelation of $f_t^\mathrm{fluct}$ occurs on time-scales $\tau$ that is small compared to time scales on which $X$ changes significantly.
If the microstates remain for a long time in or near subspaces, and these subspaces are not described well by the macroscopic variables $X$, then one might see a breakdown of fluctuation dissipation relations.

If the subspaces $A$ and $B$ are dynamically disconnected, or far apart, there is a (long) lasting dependence (of the microstate) on the initial microstate, but this might not be apparent on the macro level at all.
For typical initial states in either of the two subspaces, corresponding to the same $X$, might decorrelation in a similar way.
In this case the dependence does not show up in the correlation behavior of the fluctuations and is of no importance for the evolution at the macroscopic level.

If this is the case the (ensemble) expectation value of the correlation of $f_t^\mathrm{fluct}$ appearing in Eq.~\eqref{eq:generalized_Langevin} will be very close to the time average starting in any microstate consistent with macrostate $X_{t-s}$.
This means that fluctuations obey the fluctuation-dissipation relation.

\subsection{Consistent microstates}

For the usual many particle systems the general Birkhoff ergodicity theorem is of little use since a extremely long time $T$ is needed to sample microscopic measure $\mu_\mathrm{erg}$ the sufficiently accurate.
In general, however, one is interested in knowing average values of macroscopic observable, or dynamics of macroscopic states.
If many, possibly widely separated microstates, correspond to the same macrostate one does not necessarily need to sample the full microscopic space to sample the macroscopic space well.
Generally, for big systems, one only needs to sample a tiny amounts of microscopic phase space.
To illustrate this point let's make a digression.

As an example let's consider a box of box $V$, filled with $N$ ideal gas particles and energy $E$.
According to the entropy definition in this paper one has
\begin{equation}
  \begin{split}
    \exp[S(E, N, V)] &= \int \prod_{i=1}^N d\Gamma_i \; \delta (E - \frac{1}{2m} \sum_{j=1}^N \vec{p}_i^2)\\
    &= \frac{V^N \, (2 \pi \, m \, E)^\frac{3N}{2}}{\Gamma(3N/2)\, E}.
  \end{split}
  \label{eq:ideal_gas_entropy_1box}
\end{equation}
Note that in this expression no Planck constant occurs and no factor $1/N!$ that makes the entropy extensive.
Here $E$ and $N$ are in principle dynamical variables, but because of the conservation of energy and number of particles they will, in fact, not change.

Now consider space divided into $m$ cells with volumes $V_i$.
The macroscopic state in each cell is characterized by the energy $E_i$ and the number of particles $N_i$.
The entropy of the macroscopic system characterized by $\vec{E}=(E_1, \cdots , E_m)$, $\vec{N}=(N_1, \cdots , N_m)$, $\vec{V}=(E_1, \cdots , E_m)$.
The total entropy of the macroscopic system characterized by these variables is
\begin{equation}
  \exp[S(\vec{E}, \vec{N}, \vec{V})] = \frac{N!}{N_1! \cdots N_m! } \prod_{i=1}^m \exp[S(E_i, N_i, V_i)].
  \label{eq:ideal_gas_entropy_more_boxed}
\end{equation}
Here the multinomial factor did arise because classical particles are distinguishable!
The entropy as calculated for one cell Eq.~\eqref{eq:ideal_gas_entropy_1box} considers $N$ specific particles.

If one puts the particles labeled $1$ to $N_1$ into cell 1, labels $N_1+1, \cdots, N_1+N_2$ into cell 2 etc. and computes microscopic phase space around energy states $E-1$ to $E_m$ one finds the product of $\exp[S(E_i, N_i, V_i)]$'s in Eq.~\eqref{eq:ideal_gas_entropy_more_boxed}.
At this point one, however, only computed one out of the $N!/(N_1! \cdots N_m!)$ permutations.
To compute all microscopic phase space volume consistent with the macroscopic state one has to multiply the contribution of a specific configuration with the number of permutations.
Clearly, if a system is large, in any reasonable time only a small portion of all possible permutations will be visited by the path through microscopic phase space.
For computation of expectation values this is no problem if the macroscopic state depends on the number of particles in each cell, but not on which specific particle is where.

The reasoning why a factor $1/N_i!$ arises in Eq.~\eqref{eq:ideal_gas_entropy_more_boxed} is different from the orthodox reasoning.
Usually one reasons that it occurs because particles are indistinguishable.
The factor is then introduced in Eq.~\eqref{eq:ideal_gas_entropy_1box}, because one would otherwise be counting one microscopic state multiple times.
The current argument that all permutations of particles as distributed over the cells, consistent with the occupations, should be counted can also be found in \cite{Pen69}[chapt. V].

The conventional approach resolves the Gibbs paradox.
In classical mechanics the notion of indistinguishability is, however, not well established.
Are red and green particles distinguishable for a color blind person?
And what about properties we are currently, but possibly not in the future, blind for?
These kinds of questions on the conventional approach are addressed in \cite{Jay92}.
The approach used is that, when computing the entropy, one distinguished between red and green when the number of red and green particles in a cell are macroscopic variables.
This gives a combinatorial factor that distinguishes between colors.
If one chooses to take color into account in the macroscopic description the difference is irrelevant.
The choice of macroscopic variables determines the entropy.
Clearly, if one does not take into account macroscopic variables that are relevant this does show up as a lasting dependence on an initial microscopic state.
Of course, in practice, one chooses variables that one observes or measures in the phenomena one wants to describe.

This interpretation of the factorial factor in the entropy definition also gives a seemingly conceptual difficulty.
We accept now that particles are distinguishable.
All microscopic states corresponding to a macroscopic state are taken into account by the entropy definition.
This means that for two particles far apart, the microscopic state corresponding to the interchanging of the two will never be reached in a reasonable macroscopic time.
Nevertheless both microstates contribute to the entropy.
Note, however, that the projection operator formalism does not demand this kind of ``ergodic'' properties.
The only thing that is really important that, when interchanging the particles, fluctuations decay in a similar way.
This is for sure the case, if this interchange leaves the microscopic Hamiltonian invariant.

Another objection one might have is that one does not want to divide a system in cells.
One just wants to consider one system.
However, if the number of particles can change (without chemical reactions) one needs to model an open system.
An open system consists of 2 ``cells'' the system and the environment.

In general, to compute the entropy, one should take into account all microstates consistent with a macrostate and not only the states actually sampled.
The ergodic point of view that entropy has to do with phase space visited in a certain time is wrong.

An illustrative example for this is the entropy of a, high molecular weight, entangled polymer melt.
Upon deformation the polymer chains get stretched (on average).
Subsequently the polymer conformations will try to relax towards equilibrium.
Initially this relaxation is quick but soon polymer molecules will start feeling their neighbors.
Because the melt is entangled with them the fast modes of relaxations are halted.
According to the theory of Doi and Edwards \cite{Doi86} conformations will be confined to a tube-like region.
The contour-length and the cross sectional area of the tube is independent of the deformation.
A polymer can only relax further by escaping the tube (so-called reptation).
So, there is a two step process of relaxation, namely, a fast process of the chain inside the tube and a slow one of the tube itself.
There is a big gap between the characteristic time scales.

Here comes the point.
Suppose after a step-strain and subsequent fast relaxation inside the tube one characterizes the state by the strain.
One want to now the entropy as a function of the strain.
Since the tubes are not relaxed yet one might think the entropy can be computed from the number chain conformations (or, rather, microscopic phase space volume) sampled by a chain inside a tube.
Since the contour length and radius of the tube does not change with deformation one finds this phase space volume is independent of strain.
The mistake is that, in fact, also the number of tube configurations, consistent with the strain-deformation should be taken into account, even if they are not sampled ergodically.

Ergodic-reasoning on the number of micro-states sampled gives the incorrect result.
The correct result is found when considering all states consistent with the macroscopic description.

\subsection{Extensivity}

For systems that are totally independent the entropy is an additive quantity, since the volume of phase space corresponding to the total system is the volumes of the individual systems multiplied.
For systems that interact weakly this is still the case, if the macroscopic quantities are quantities of these subsystems, say the energy of the systems.

A special situation occurs when the systems can interchange particles, because the possible microstates consistent with an occupation numbers, $N_i$, of cells in space increases by an immense factor due to all possible permutations, see Eq.~\eqref{eq:ideal_gas_entropy_more_boxed}.
The additivity rule is maintained when using $S(E_i,N_i,V_i)- \ln N_i!$.

We will discuss the case where the variable is total energy, but it is valid for any (conserved or non-conserved) extensive additive quantity.
If one wants to know the entropy as function of the total energy of all subsystems combined the systems are no longer independent.
The energy that leaves one subsystem has to enter another one.
The total entropy for this situation is computed in appendix \ref{sec:routes}.
The total microscopic phase space per unit total energy is given by Eq.~\eqref{eq:total-entropy5}.
The entropy to leading order in the number of subsystems $M$ is then
\begin{equation}
    S(E_\mathrm{tot}) =M \, \biggl [ S(\bar{E}) - \frac{1}{2} \ln \Bigl( \frac{- S''(\bar{E})}{2\pi} \Bigr) \biggr] + o(M),
    \label{eq:extensive_entropy}
\end{equation}
where $S(\bar{E})$ is the entropy of a subsystem at mean energy $\bar{E}=E_\mathrm{tot}/M$.
The small o-notation is used to indicate weaker than linear terms.
Possibly surprising for some is that even in the leading order of $M$ a second term, besides $S(\bar{E})$, is present.
This term is negligible if the subsystem itself is already macroscopic, but otherwise not.
It is instructive to consider the case of the ideal gas entropy dependence on $E$.
For a subsystems with $N$ particles, where according to Eq.~\eqref{eq:ideal_gas_entropy_1box}, $S(E)= (3 \, N-2)/2 \ln E$ (plus constants independent of $E$).
Inserting this relation into Eq.~\eqref{eq:extensive_entropy} gives
\begin{equation}
  \begin{split}
    S(E_\mathrm{tot}) &= \frac{M \, (3 \, N-2)}{2} \ln E - \frac{M}{2} \ln \Bigl( \frac{3 \, N-2} {4 \pi\, E^2} \Bigr)\\
    &= \frac{3\, M  \, N}{2} \ln E_\mathrm{tot} + \cdots
    \label{eq:ideal_gas-limit}
  \end{split}
\end{equation}
Here the terms that are left out are independent of the energy.
If one would assume that the total entropy is simple $M \,  S(\bar{E})$ one would only find the first term in the first line of Eq.~\eqref{eq:ideal_gas-limit}.
If a subsystem with a small number of particles, $N$, was chosen then, in the limit of large $M$, one would find the wrong value.

When describing macroscopic systems one usually makes use of densities.
For large enough, homogeneous, systems entropy becomes to a good approximation extensive,
\begin{equation}
  S(X) = M \, s(X/M) + o(M).
  \label{eq:extensive-entropy}
\end{equation}
Here the entropy density, as discussed above, can not be exactly identified with the entropy of a subsystem.
The conceptual approach to macroscopic equations is to consider macroscopically large spatial ``cells''.
These volumes contain a large number of weakly interacting subsystems.
If macroscopic quantities change little from one cell to the next, one can approximately describe this by continuously varying (density) fields.

The primary quantities, however, are still the values of total (or averaged) quantities inside the underlying cells of macroscopic magnitude.
If $X$ is the macroscopic quantity under consideration in the cell, one can define a local density $x=X/M$.
According to the coordinate transformation rule, Eq.~\eqref{eq:coordinate_transform}, combined with Eq.~\eqref{eq:extensive-entropy},
\begin{equation}
  S_\mathrm{cell}(x) = S_\mathrm{cell}(X) + \ln M = M \, s(x) + o(M).
\end{equation}
The important thing to notice is that here still, $M$ is present.
The reason is that although one considers a density it characterizes the cell consisting of $M$ subsystems.
Contrary to common believe in the transition to densities one does not lose the (sub)system size dependence.
Often, when treating the thermodynamic limit, e.g., in large deviation theory, one calls $s(x)$ instead of $S(X)$, the entropy.

If system size $M$ is large enough, it can be thought of consisting of independent (weakly dependent) subsystems.
The $M$ subsystems that constitute a cell fluctuate independently.
Therefore, the fluctuations for, e.g., a density $x=X_\mathrm{tot}/M$ scales as,
\begin{equation}
  D \propto \frac{1}{M}.
  \label{eq:fluctuation_macroscopic}
\end{equation}
If one considers Eq.~\eqref{eq:SDE} the product $\ten{D}\cdot \partial S/ \partial x$ will become independent of $M$.
The divergence term of $D$ and the fluctuating term will vanish for $M \rightarrow \infty$.  
In the case of macroscopic equations one implicitly uses the following reasoning.
First one assumes that quantities vary slowly such that one can consider a ``discretization'' in macroscopically large cells.
Because $M$ is large the fluctuating term (and the divergence term) in the stochastic differential equation can be neglected and one obtains a ordinary differential equation.
Because this equation is independent of the cell size (no $M$ dependence) one can introduce smooth fields and take $M \rightarrow 0$.
Although one is now considering infinitesimal small cells one must keep in mind the route taken to arrive here.

The interest of the author is mainly in simulations on a mesoscopic level.
For many applications on the macroscopic level one can assume an extensive entropy.
When considering smaller scales fluctuations will start to play a role.
Going from the macroscopic level to the mesoscopic one, initially fluctuations will scale as Eq.~\eqref{eq:fluctuation_macroscopic} and the entropy can still taken to be extensive.
At still smaller scales the extensivity breaks down.
Here one needs to resort to the microcanonical definition of entropy, or at least take non-extensive contributions (such as interfacial interaction) into account.

\subsection{Other entropy expressions}

It is well accepted that the microcanonical ensemble is more elementary than the macrocanonical one.
There is more debate on the entropy definitions.
Some researchers believe the microcanonical definition is elementary, others think microcanonical and macrocanonical are on the same footing (and therefore entropy is only well defined in the thermodynamic limit), still others have a preference for Gibbs entropy (or information theoretical Shannon-Jaynes entropy).

To a certain extend this seems a matter of personal preference, since ensembles are equivalent.
This equivalence is, however, only the case when the thermodynamic limit is valid.
Even then, equivalence can only be established if certain requirements are obeyed.
Mainly interactions have to be short ranged with a finite attractive part.
The equivalence of ensembles in the thermodynamic limit, and the requirements the potential  has to obey, is well established \cite{Rue69, Lan73}.

The modern approach to the equivalence of ensembles in the thermodynamic limit is the theory of large deviations \cite{Ell99}.
In unsophisticated terms the basic premises is that (for short range interactions), large systems can be divided into subsystems that interact only weakly.
In this case one can do statistics and count the number of systems that are in a certain state.

One of the candidates for the for the fundamental definition of (thermodynamic) entropy is the Gibbs entropy, i.e., the integral of $-\rho \log \rho$.
For example, in the information theoretic (MaxEnt) approach to classical mechanics, \cite{Jay89}, uses this entropy as starting point.
The case for the information-theoretic entropy definition on the basis of a measure of uncertainty, \cite{Sha48} are quite strong.
The information entropy is often defined in a axiomatic way, \cite{Sha48, Sho80}.
Implicit in the definition, \cite{Sha48}, is that entropy is additive.
This can be shown to follows from the fact that entropy maximizes uncertainty constraint by prior information, \cite{Sho80}.
The main assumption is, therefore, the Baysian probability interpretation combined with a maximization procedure.

In the case of continuous distributions only relative (Gibbs) entropy, i.e., with respect to a specified measure, can be rigorously defined.
Therefore one first needs to establish the origin of this measure.
In classical mechanics this measure is the Liouville measure.
So, prior to be able to use the Gibbs or Shannon-Jaynes entropy one needs to argue why this measure can be used by, e.g., a reasoning based on ergodicity.

If one uses a rigid constraint (such as total energy fixed) one finds from the MaxEnt approach the microcanonical ensemble\footnote{Although in this case, because of its localized nature, the information-entropy itself is not finite.}.
Therefore the microcanonical entropy is sometimes said to be a special case of the Gibbs entropy.
This is not such a strong argument because already the Liouville measure and the constraint to the iso-energy shell are put in as ingredients.
Usually the information-entropy is maximized using expectation values, such as average energies, as constraints.
When using an expectation value as constraint for a conserved quantity as the energy one implicitly states that the system is open.

When trying to define the Gibbs entropy in more physical terms one, inevitably, ends up with deriving information entropy as a limit of a multinomial distribution. (Even Jaynes does this, see \cite{Jay03}[p. 351])
In appendix \S\ref{sect:rel-entropy} such a derivation is given.
The main assumption is that there are many weakly interacting subsystems.
The density of states $\rho$ should be interpreted as counting the number of subsystems in a certain state.
This is also the way relative entropy appears in the theory of large deviations \cite{Ell99}.
The total entropy for, e.g., constrained total energy, can be computed by a functional integration over all possible ensembles, $\rho$.
The maximization principle arises because of a saddle-point approximation to this integral.

Defining the Gibbs entropy for the instantaneous phase-space distribution $\rho(\Gamma)$ for an isolated system is without foundation.
Therefore the well known fact that, as a consequence of Liouville's theorem, the Gibbs entropy is a constant of motion poses no paradox.
Sometimes a coarse-grained Gibbs entropy is defined to circumvent this perceived paradox.
Instead of the distribution $\rho$, following from the classical Liouville equation, a smoothed one $\bar{\rho}$ is used.
The motivation for this might be a quantum mechanical reasoning that volume of elementary cells need to be larger than $\hbar^{3N}$.
The smoothing causes diffusion in phase space which gives rise to an increase in Gibbs entropy.
This coarse graining procedure is not well defined and rather ad-hoc.
A critique on the approach can be found in, e.g., \cite{Zub74}.

Note that often in entropy expressions in, e.g., mean-field theories some terms are called Gibbs-entropies which are strictly speaking not.
Instead of a density that counts the number of subsystems there is a density that counts the number of particles in a region in space.
In that case $-\rho \log \rho$ arises from the approximation of the multinomial in Eq.~\eqref{eq:ideal_gas_entropy_more_boxed}\footnote{Here one should take care to exclude $- \ln N_i!$ in the a local entropy definition.}.
This kind of entropy, that really is just an ideal gas entropy, is also used in the $H$-entropy defined for the Boltzmann equation for dilute gasses.

We conclude that the Gibbs-entropy is a very useful tool for use in extensive systems.
It does not provide a fundamental definition of entropy, but follows from it if certain requirements are obeyed.
If systems are to small, clearly not extensive, fluctuations play a large role and the maximization procedure is not a good approximation.

In the thermodynamic limit, under the considerations as given by \cite{Rue69}, the thermodynamic entropy always will be a concave function of energy.
For finite systems there might be a ``convex'' intruder.
In this case the total entropy of system plus heat-bath will be bimodal for some temperature interval.
In this case there is no complete equivalence between the microcanonical ensemble and the macrocanonical ensemble, \cite{Gro05, Ell00}.
This equivalence can break down completely if interactions are long ranged such as in the case of gravity.
In this case one needs to resort to the microcanonical ensemble itself, or at least to approximations better than the canonical ensemble (e.g., the Gaussian or generalized canonical ensemble \cite{Cos05})

Another ``fundamental'' entropy definition based on the microcanonical ensemble can be found in literature, e.g., in \cite{Umi99, Dun06}.
The textbook references are \cite{Bec67, Mun69}
Here, the ordinary thermodynamics entropy is defined as the logarithm of all phase space volume for states with energies smaller than $E$,
\begin{equation}
  \exp[S(E,V)] = \int d\Gamma \, \Theta[E - H(\Gamma; V)],
  \label{eq:Becker-entropy}
\end{equation}
where $\Theta[\;]$ is the Heaviside step-function.
The main reason why one would prefer this definition is that, also for finite systems,
\begin{equation}
  dS = \frac{1}{T} dE -\frac{p}{T} dV,
  \label{eq:thermodynamic}
\end{equation}
For the entropy based on the microcanonical ensemble, Eq.~\eqref{eq:entropy_definition2}, the relation only holds in the thermodynamic limit (see \cite{Rue99} for the equivalence in the thermodynamic limit).
The thermodynamic relation Eq.~\eqref{eq:thermodynamic} assumes entropy is an additive quantity.
Since this is only rigorously valid in the thermodynamic limit the objection that Eq.~\eqref{eq:entropy_definition2} does not obey Eq.~\eqref{eq:thermodynamic} is not very serious.
Moreover it is hard to see why states with energies that are not attained by the system should be important for dynamics.
Moreover generalizations of Eq.~\eqref{eq:Becker-entropy} for other variables than $E$ give troubles if this variable is not bounded from below, such as the energy.

In his work on small systems, \cite{Hill63}, Hill discusses the thermodynamics of small systems.
Clearly, extensivity is not valid and therefore the Gibbs-Duhem relation breaks down.
As a tool he introduces a new variable, $\mathcal{N}$, the number of identical small systems.
Identical means that the small systems are all characterized a the same set of the``extensive'' variables (e.g., energy) and intensive variables (e.g., chemical potential and pressure). 
For small systems there is a difference if variables, such as energy are fixed, or allowed to fluctuate.
In his treatment this is, however, not the case.
He asserts that still thermodynamic relations such as, Eq.~\eqref{eq:thermodynamic}, are valid.
Therefore in his treatment, there is not a difference between systems characterized by $\bar{E}=E_t/\mathcal{N}$, (i.e., total energy divided by the number of independent ensembles), and non fluctuating energy $E$, which is the same each small systems in the ensemble of $\mathcal{N}$ members.

In a digression on statistical mechanics Hill uses a Gibbs-entropy definition.
As discussed above, for sufficiently small systems, taking the entropy to be equal to the maximum of Gibbs entropy, is not a valid assumption.
This can be defended for open systems if $\mathcal{N}$ is large.
In one case the total entropy, $S_t$, for an ensemble of, e.g., systems with all the same energies, $E$, is computed.
Compared to an entropy that depends on $\bar{E}$, many possible states, such as system 1 with energy $E_1$ and system 2 with energy $E_2$, etc., are left out of consideration.
Nevertheless, it is implicitly assumed that both entropies are the same.

\section{Conclusions}

In this paper we showed that a rigorous definition of entropy follows from the derivation of generalized Langevin equation using projection operator formalism.
This is a purely formal procedure.
The only physical input to obtain Eq.~\eqref{eq:generalized_Langevin} is Liouville's theorem.
The entropy definition is close to the Boltzmann definition.
The subtle difference is that the exponent of the entropy is not the number of states per macrostate, but the volume of microstates per unit macrostate-volume.
Entropy can be fully defined within a classical mechanics framework without the appearance of any paradoxes that need quantum mechanical reasoning to resolve.

The entropy definition follows from a projection onto coarse-grained variables of the Liouville equation describing the dynamics of the system.
No equilibrium, or ergodic, reasoning is used to define entropy.
There is a straight, deductive, route from microscopic dynamics to (non-)equilibrium thermodynamics.
Entropy is in some sense subjective since it depends on the choice of variables to describe a system.
When one speaks about {\em the} entropy in the setting of equilibrium thermodynamics one means a very specific one, namely the entropy as function of energy and volume of a system (that interacts weakly with its environment).
For describing different phenomena one can choose to compute a entropy as function of different quantities.
Entropy has also a objective quality since it refers to microscopic phase space volume in a well defined way once the macroscopic variables are fixed.

The notion of entropy is always related to dynamic variables.
The reason one wants to know entropy, e.g., as a function of energy, is that one wants to be able to make predictions about heat-fluxes, i.e., change of energy.
It makes no sense to discuss total entropy of a closed system as function of total energy, for example in the case of the universe.
Entropy only becomes a useful notion if one divides the system into subsystems and characterizing each subsystem by macroscopic variables.

Entropy as defined in this paper is not a scalar quantity.
So upon a change of variables extra terms appear.
In the thermodynamic limit these terms are negligible.
It has, however, consequences for small systems.
In this case the current entropy definition deviates from other ones such as the Gibbs entropy.
Because of the rigorous connection through Zwanzig projection operator formalism with microscopic dynamics the current entropy definition is proved to be the correct one to use.
Moreover, if one approximates the governing equation with a stochastic differential equation the transformation rule, Eq.~\eqref{eq:coordinate_transform}, is essential.
Only when allowing the entropy to transform in this way the form of the equation does not change upon a change of coordinates, as follows from Ito-calculus.

In the thermodynamic limit a Gibbs entropy definition can be deduced from the more fundamental entropy definition given here.
This Gibbs entropy implies a MaxEnt procedure to compute coarse(r) grained entropies from the Gibbs entropy.
The procedure consists of two steps, one is determine a (constrained) maximum of the Gibbs entropy.
The second step is integration around this saddle-point (in the complex plane).
The integral gives a factor whose variation is negligible only in the thermodynamic limit.

The Langevin equation poses no restriction on the set of variables one uses to describe a system.
The choice should be motivated by the problem at hand.
What determines a good choice is decorrelation behavior of fluctuations of the macroscopic variables.
If they decorrelate quickly the formal generalized Langevin equation can be approximated by a practically useful (stochastic) equations.

The current entropy definition is independent of (local) equilibrium assumptions.
It is therefore suited for non-equilibrium modeling.
One should not put too much emphasis on ergodicity reasoning.
If fluctuations decorrelate quickly and in such a way that fluctuation-dissipation relations are found to be obeyed, then it is no problem that only a small part of the microscopic phase space corresponding to a macroscopic state is sampled.
It can easily be shown that many microscopic states that contribute to the entropy are not approached, even remotely, even if one waits a very long time.
Here the notion of an equivalence class, i.e., fluctuations behave similarly in this remote corner of phase space, is important.

Entropy does not measure the (logarithm of) states sampled.
It measures all phase space volume consistent with a macroscopic space.
This means that much of this microscopic phase space that are accounted for in the entropy might actually not be sampled.
In the Langevin equation differences of entropy are the driving force.
This corresponds to a ratio of volumes.
The fact that a part of phase-space is not sampled is not important if the motion in the (dynamically) disjointed regions are typical or equivalent.
The ratio denotes how much phase space opens up if the (macro)system evolves in a certain direction.
If this ratio is the same for all equivalent disjointed regions it's okay.

Only taking into account of phase space, in some way, sampled in a characteristic time can lead to erroneous results.
The number of disjointed regions (if this can be defined at all), might depend on the macroscopic state.
Regions can therefore open up to multiple regions, or regions might merge upon a change of the macroscopic state.

The most common approximate modeling for fluctuations is to describe them as white noise.
The equation that arises is very close to those in the GENERIC formalism \cite{Ott05}, but slightly more general.
One difference is that the entropy is more rigorously defined in the current paper.
GENERIC also imposes a more strict structure on the reversible part of the stochastic differential equation that arises.
The reversible part has the energy as a ``generator''.
To arrive at this result one needs to make extra approximations, e.g., introduce an internal energy.
Sometimes these approximations are hard to make for the macroscopic variables chosen, e.g., in the case of a Brownian particle.
Also the assumed Poisson structure of the reversible part remains unproven.
The GENERIC structure also seems to miss a possible anti-symmetric part driven by entropy differences that can arise due to Casimir anti-symmetries.
The degeneracy conditions that are assumed to hold in the GENERIC framework where proved to follow from the properties of the generalized non-linear Langevin equation.

The approach used in this paper agrees with the ``typicality'' point of view in \cite{Leb93,Leb99}.
Upon the coarse-graining the equations of motion generate typical paths through the macroscopic phase space.
The entropy needed is the Boltzmann-Planck entropy since it quantifies this typicality.
Motion toward states corresponding to a larger microscopic phase space volume (Liouville measure) are biased because a microscopic path will typically move in this direction.

In conclusion, the entropy definition deduced in this paper is a definition that is generally valid also outside the thermodynamic limit and in far from equilibrium situations.

\appendix

\section{Routes to an extensive entropy}
\label{sec:routes}

Let's consider a system divided into many identical subsystems.
The subsystems interacts so weakly that, to a good approximation, the total energy can be written as a sum of energies.
The goal of this appendix is to compute ``the entropy'' of such a system. 
We will present only an outline.
For (mathematical and physical) subtleties see \cite{Rue69, Lan73}. 
The purpose is to show the entropy definition as discussed in this paper in action and show interconnections with alternative entropy expression (which are strictly valid only in the thermodynamic limit).

The systems evolve almost independently (not fully because they exchange energy).
Let $\Gamma = (\Gamma_1, \cdots, \Gamma_M)$ be the microscopic state of the full system and $\Gamma_j$ the microstates of the subsystems.
The subsystems will interact with their neighbors but this interaction is weak.
Therefore to a good approximation the total energy is
\begin{equation}
  E_\mathrm{tot}(\Gamma) = \sum_{i=1}^M E(\Gamma_j).
\end{equation}
We want to compute the entropy of the total system as function of the total energy, i.e.,
\begin{equation}
  \begin{split}
    \exp[S(E_\mathrm{tot})] &= \int d\Gamma \, \delta[E(\Gamma)-E_\mathrm{tot}]\\
    &= \int \prod_j d\Gamma_{j=1}^M \, \delta[\sum E(\Gamma_j)-E_\mathrm{tot}].
  \end{split}
  \label{eq:total-entropy}
\end{equation}
Next we introduce a Fourier representation for the Dirac delta-function,
\begin{multline}
    \exp[S(E_\mathrm{tot})]= \\
    \frac{1}{2\pi} \int_{-\infty}^\infty \!\! dk \int \prod_j d\Gamma_j \, \exp[i\, k \, \sum E(\Gamma_j)- i\, k\, E_\mathrm{tot}]\\
    = \frac{1}{2\pi} \int_{-\infty}^\infty \!\! dk \left ( \int d\tilde{\Gamma} \, \exp[i\, k \, E(\tilde{\Gamma})] \right )^M \exp[- i\, k\, E_\mathrm{tot}].
  \label{eq:total-entropy2}
\end{multline}
Let us introduce a sum of states
\begin{equation}
  Z(\beta) = \int d\tilde{\Gamma} \, \exp[- \beta \, E(\tilde{\Gamma})],
  \label{eq:sum_of_states}
\end{equation}
here $\beta$ can be a complex number.
For finite systems, with $E(\tilde{\Gamma})$ well behaved $Z(\beta)$ is analytic everywhere on $\mathrm{Re}(\beta)>0$.
Besides this $Z(\beta)$ for real $\beta>0$ will be always positive.
Assuming that $Z(\beta)$ is analytic on the plane $\mathrm{Re}(\beta) \ge 0$ and on this plane decays rapid enough when $|\beta| \rightarrow \infty$ one can change the path of integration from running along the imaginary axis to run along $\beta - i\,k$
Using definition Eq.~\eqref{eq:sum_of_states} in Eq.~\eqref{eq:total-entropy} gives
\begin{multline}
    \exp[S(E_\mathrm{tot})]= \\
    \frac{1}{2\pi} \int_{-\infty}^\infty \!\! dk \int \prod_j d\Gamma_j \, \exp[i\, k \, \sum E(\Gamma_j)- i\, k\, E_\mathrm{tot}]\\
    = \frac{1}{2\pi} \int_{-\infty}^\infty \!\! dk \exp[M \, \ln Z(\beta - i\,k) + (\beta - i\, k) \, E_\mathrm{tot}].
  \label{eq:total-entropy3}
\end{multline}
Here $\beta$ is still free to choose.
A particular convenient choice is,
\begin{equation}
  \frac{d}{d\beta}(M \, \ln Z(\beta) + \beta E_\mathrm{tot})=0.
  \label{eq:saddle}
\end{equation}
This is a saddle-point condition for the term in the exponent of Eq.~\eqref{eq:total-entropy3}.
The $\beta_\mathrm{sad}$ thus found is the inverse temperature.
One can perform a Taylor expansion up to second order around the saddle point.
This gives
\begin{equation}
  \begin{split}
    \exp[S(E)] & \approx
    \exp[M \, \ln Z(\beta_\mathrm{sad}) + \beta_\mathrm{sad} E_\mathrm{tot}] \\
    & \quad \times \frac{1}{2\pi} \int_{-\infty}^\infty \! \! dk \exp\Bigl[ -\frac{M}{2} k^2 \frac{d^2}{d \beta^2} \ln Z(\beta_\mathrm{sad}) \Bigr ]\\
    &= \exp[M \, \ln Z(\beta_\mathrm{sad}) + \beta_\mathrm{sad} E_\mathrm{tot}]\\
    & \quad \times \Bigl (2 \pi M \frac{d^2}{d\beta^2} \ln Z \Bigr)^{-\frac{1}{2}}. 
  \end{split}
\label{eq:total-entropy4}
\end{equation}

At a few points in the computation we could have made the decision to first compute the entropy for the subsystems,
\begin{equation}
  \exp[S(E)] = \int d\tilde{\Gamma} \, \delta[E(\tilde{\Gamma})-E],
\end{equation}
then
\begin{equation}
    \exp[S(E_\mathrm{tot})] \approx \int \prod_j dE_j \, \exp\Bigl[\sum_j S(E_j)\Bigr] \, \delta\Bigl[\sum E_j-E_\mathrm{tot} \Bigr] .
\end{equation}
Taylor expanding $S(E_j)$ around the average energy $\bar{E}=E_\mathrm{tot}/M$ up to second order gives
\begin{multline}
  \exp[S(E_\mathrm{tot})] \approx \exp [M \, S(\bar{E})] \int \prod_j dE_j \\
   \quad \times  \exp\Bigl[\frac{1}{2} S''(\bar{E}) \sum_j (E_j-\bar{E})^2 \Bigr ] \, \delta\Bigl[\sum (E_j-\bar{E}) \Bigr]\\
      = \exp [M \, S(\bar{E})] M^{-\frac{1}{2}} \Bigl( \frac{-2\pi}{S''(\bar{E})} \Bigr)^\frac{M-1}{2}.
      \label{eq:total-entropy5}
\end{multline}

\section{Relative Entropy}
\label{sect:rel-entropy}

The notion of relative entropy, arises naturally if one considers a large ensemble of independent (sub)systems.
Let's consider $M$ subsystems, where each of the systems can be in any one of the discrete states $i$ with measure $\nu_i$.
Since the systems are independent one can define a product measure (hyper-volume).
The measure of system 1 to be in state $i_1$, system 2 in state $i_2$, etc. for $M$ systems is
\begin{equation}
  \nu^\mathrm{(prod)}_{i_1,i_2,\dots, i_M} = \prod_{j=1}^M \nu_{i_j}.
\end{equation}

Alternatively one can count the number of system that are all in state $i$.
Let this number be $M_i$.
Of course $\sum_i M_i =M$.
So, the total product measure of having $M_1$ subsystems in state $1$, $M_2$ in $2$ etc.,
\begin{equation}
  \begin{split}
    \nu^\mathrm{(prod)}(M_1, \cdots, M_n) &= \frac{M!}{M_1! \cdots M_n!} \prod_{i=1}^n (\nu_i)^{M_i}\\ &= \exp[ - M \sum_i \rho_i \ln (\rho_i/\nu_i) + o(M)],
  \end{split}
  \label{eq:product_measure}
\end{equation}
where $\rho_i=M_i/M$.
Here the Stirling approximation is used for the factorials in the multinomial.
If one wants to compute the total entropy, say as function of the total energy $E_\mathrm{tot}$, one finds
\begin{equation}
  \begin{split}
    \exp[S(E_\mathrm{tot})] & =  \sum_{\sum_i M_j = M} \delta[E - \sum_i M_i E_i] \\  
    & \quad \times \nu^\mathrm{(prod)}(M_1, \cdots, M_n)\\
    &\approx \sum_{\sum_i M_j = M} \delta[E - \sum_i M_i E_i] \\
    &\quad \times  \exp[ M S_\mathrm{rel}(\rho_1, \cdots, \rho_n) + o(M)]
  \end{split}
  \label{eq:total_entropy_6}    
\end{equation}
Here the quantity,
\begin{equation}
  S_\mathrm{rel}(\rho_1, \cdots, \rho_n) = - \sum_i \rho_i \ln (\rho_i/\nu_i),
\end{equation}
is called the relative entropy.
The relative entropy gives, to leading order, the exponent in Eq.~\eqref{eq:total_entropy_6}.
Maximization of the relative entropy, subject to certain constraints, such as $E_\mathrm{tot}/M = \bar{E}= \int \rho(d \Gamma) \, E(\Gamma)$, can be used in for a saddle-point approximation of Eq.~\eqref{eq:total_entropy_6}.

The relative entropy can be straightforwardly generalized to the continuous case.
If the state are taken to be macrostates the measure will be the measure of the underlying microscopic space, i.e., $\exp[S(X)] \mu_L(dX)$, so
\begin{equation}
  S_\mathrm{rel}(\rho) = - \int \rho(dX) \ln (\rho(dX)/\mu_L(dX)) + \int \rho(dX) \, S(X).
  \label{eq:rel_entropy}
\end{equation}
In this point of view, Gibbs-entropy is a special case of relative entropy.
For the microscopic space entropy is zero, the measure is the Liouville measure, i.e., $\mu_L(d\Gamma)$.

Note that from the viewpoint discussed here $\rho(d\Gamma)=\rho(\Gamma) \, d\Gamma$ is proportional to the number of systems near microstate $\Gamma$.
The subsystems are not isolated.
They can, in the example above, exchange energy.
Therefore defining a Gibbs-entropy for a phase space density, $\rho(\Gamma)$, of the ensemble of an isolated system evolving exactly according to a fixed Hamiltonian, can not be justified.

Relative entropy always arises as an approximation of a multinomial.
In the derivation of the extensive entropy in appendix \ref{sec:routes} one could have made a change of variables and use relative entropy.
If on writes the integral over $\tilde{\Gamma}$ in Eq.~\eqref{eq:total-entropy2} as a Riemann sum one finds
\begin{multline}
   \left ( \int d\tilde{\Gamma} \, \exp[i\, k \, E(\tilde{\Gamma})] \right )^M \\ \approx \left ( \sum_p \Delta\tilde{\Gamma} \, \exp[i\, k \, E(\tilde{\Gamma}_p)] \right )^M \\
= (\Delta\tilde{\Gamma})^M \sum_{\sum M_p = M} \left( \frac{M!}{M_1! \cdots  M_n!} \exp[i\, k \, \sum_p M_p E(\tilde{\Gamma}_p)] \right ).
\end{multline}
Now one can compute an entropy as function of the occupations $M_p$, i.e., the number of subsystems in a state near $\tilde{\Gamma}_p$.
The multinomial can be approximated using the relative entropy.
One can perform a saddle-point approximation to compute the entropy as function of the total energy.
This involves maximization of the relative entropy with a constraint on the average energy.

\begin{acknowledgments}
The research of Dr.\ Peters has been made possible by a fellowship of the Royal Netherlands Academy of Arts and Sciences.
\end{acknowledgments}

%\bibliography{../frank}

\end{document}